\documentclass[aps,pra,twocolumn,showpacs,notitlepage,nofootinbib]{revtex4-1}
\usepackage{graphicx}
\usepackage{amsmath,amsfonts}
\usepackage{amssymb}
\usepackage{color}
\usepackage[T1]{fontenc}
\usepackage{hyperref}
\usepackage{color}
\begin{document}
\graphicspath{{./figures/}}
\title{Non-adiabatic quantum phase transition in a trapped spinor condensate}
\author{Tomasz \'Swis\l{}ocki,$^{1}$ Emilia Witkowska$\,^{2}$ and Micha\l{} Matuszewski$\,^{2}$}
\affiliation{$^{1}$Faculty of Applied Informatics and Mathematics, Warsaw University of Life Sciences, ul. Nowoursynowska 159, 02-786 Warsaw, Poland\\
$^{2}$Institute of Physics, PAS, Aleja Lotnikow 32/46, 02-668 Warsaw, Poland}
\begin{abstract}
We study the effect of an external harmonic trapping potential on an outcome of the non-adiabatic quantum phase transition from an antiferromagnetic to a phase-separated state in a spin-1 atomic condensate.
Previously, we demonstrated that the dynamics of an untrapped system exhibits double universality with two different scaling laws appearing due to conservation of magnetization. 
We show that in the presence of a trap double universality persists. 
However, the corresponding scaling exponents are strongly modified by transfer of local magnetization across the system. 
The values of these exponents cannot be explained by the effect of causality alone, as in the spinless case.
We derive the appropriate scaling laws based on a slow diffusive-drift relaxation process in the local density approximation.
\end{abstract}

\maketitle

\section{Introduction}

The Kibble-\.Zurek (KZ) mechanism refers to the dynamics of spontaneous symmetry breaking which takes place near a non-adiabatic second order phase transition. When the system approaches a critical point, the correlation length can no longer adiabatically follow its diverging equilibrium value. Consequently, at the critical point the transition occurs without correlation length having reached the size of the whole system. As a result, finite-sized phase domains are formed, which display independent choices of the symmetry breaking order parameter. The KZ physics have been intensively studied experimentally in a wide range of systems including experiments in atomic Bose-Einstein condensates \cite{Sadler2006, Weiler2008, White2011, Lamporesi2013, Corman2014, Brauna2014, Chomaz2015, Navon2015, Chapman2016} which indeed provide the ability of adapting and tuning the system parameters with an exceptional level of control. 

The KZ theory predicts a universal scaling law for an average domain size in terms of critical exponents characteristic for a universality class of the system. A simple theoretical framework is based on the adiabatic-impulse-adiabatic approximation and does not take into account specific processes which can lead to the change of the scaling exponent. This is the case for the system studied in this paper. We consider the quantum phase transition from an antiferromagnetic to a phase-separated state by increasing an external magnetic field. In our previous work \cite{ourPRL, ourPRB, ourPRA} we demonstrated the modification of the KZ mechanism due to conservation of a magnetization in the system. Initially, the quantum phase transition develops in usual way. The number of spin domain seeds that appear is well described by the KZ theory. The modification takes place during the growth of spin domain seeds into stable spin domains. Only some of the domain seeds develop into stable domains because of the conservation of magnetization, which determines the volume fraction of the new phase and limits the density of spin domains. This so called post-selection process results in the second scaling law with a different exponent for the number of domains in the final stable configuration. 

In this paper we extend our further analysis by considering the effect of an external harmonic trapping potential on an outcome of the quantum phase transition in this system. We find that the inhomogeneity, arising as a result of the external trapping potential, brings in new physics. Due to the spatial dependence of the critical magnetic field, different parts of the system undergo phase transition at different times. 
The transition starts in the trap center, and formation of spin domains is governed by causality. 
Previously, it was shown that in trapped systems the exponents in the KZ theory are modified by the finite velocity of the phase transittion front~\cite{Zurek2009, Dziarmaga2010, Sabbatini2012, delCampo2013, Saito2013, delCampo2014}. When the front of the transition moves faster than the characteristic velocity of perturbation, domains nucleate; otherwise the choice of the order parameter in the broken symmetry phase is done homogeneously across the system. The causality introduces a characteristic length scale of the region in which domains can nucleate, and sets values of the scaling exponents. 
We show that the outcome of the quantum phase transition in our system cannot be explained by the causality alone. In addition to the causality effect, a process of transport of local magnetization from the trap center to its remote parts occurs, leading to a new characteristic length scale, namely the size of the low-magnetization region in the trap center in which domains can nucleate. The new length scale depends on the quench rate, introducing further modifications of the two scaling exponents. We explain the corresponding scaling laws by considering a slow diffusive-drift relaxation process, obtaining excellent agreement in a wide range of quench times.

\section{Antiferromagnetic spinor condensates in 1D}

We consider a spin-1 Bose gas in one spatial dimension confined by a harmonic trapping potential, and in a homogeneous magnetic field $B$. The model Hamiltonian of the system consists of two terms. The first (spin-independent) part is
\begin{equation} \label{En}
H_0 = \sum_{m_{f}=-1,0,1} \int d x \, \psi_{m_{f}}^\dagger \left(-\frac{\hbar^2}{2m}\nabla^{2} + V(x) + \frac{c_0}{2} \rho \right) \psi_{m_{f}},
\end{equation}
where the subscripts ${m_{f}}=-1,0,1$ denote sublevels with the corresponding magnetic quantum numbers along the magnetic field axis,
$m$ is the atomic mass, $\rho=\sum \rho_{m_{f}} = \sum \psi_{m_{f}}^\dagger \psi_{m_{f}}$ is the total atom density, and 
$V(x)=m \omega^2 x^2/2$ is the external potential. 
The second (spin-dependent) part can be written as
\begin{equation} \label{EA}
H_{\rm A} = \int d x \, \left( \sum_{m_{f}} E_{m_{f}} \rho_{m_{f}} + \frac{c_2}{2} :{\bf F}^2:\right)\,,
\end{equation}
where $E_{m_{f}}$ are the Zeeman energy levels, the spin density is 
${\bf F}=(\psi^{\dagger}f_x\psi,\psi^{\dagger}f_y\psi,\psi^{\dagger}f_z\psi)$,
where $f_{x,y,z}$ are the spin-1 matrices, $\psi^T =(\psi_1,\psi_0,\psi_{-1})$ and dots in the last term denote normal ordering.
The spin-independent and spin-dependent interaction coefficients are given by 
$c_0=2 \hbar \omega_\perp (2 a_2 + a_0)/3$ and $c_2= 2 \hbar \omega_\perp (a_2 - a_0)/3$, 
where $a_S$ is the s-wave scattering length for colliding atoms with total spin $S$, and $\omega_\perp$ is the frequency of a transverse potential.
Both $c_0$ and $c_2$ are positive, ensuring the antiferromagnetic ground state \cite{Zhang2003}.
In the following analytic calculations we often assume the incompressible regime where $c_0~\gg~c_2~$, which is satisfied by e.g. a $^{23}$Na spin-1 condensate.
The total atom number $N = \int \rho d x$ and the magnetization $M=\int \left(\rho_+ - \rho_-\right) d x$ are conserved quantities.

The linear part of the Zeeman shifts $E_j$ induces homogeneous rotation of the spin vector around the direction of the magnetic field.
Since the Hamiltonian is invariant with respect to such spin rotations, 
we consider only the effects of the quadratic Zeeman shift~\cite{Matuszewski_AF,Matuszewski_PS}.
For sufficiently weak magnetic field we can approximate it by a positive energy shift of the $m_f=\pm 1$ sublevels 
$\delta=(E_+ + E_- - 2E_0)/2 \approx B^2 A$, 
where $B$ is the magnetic field strength and $A=(g_I + g_J)^2 \mu_B^2/16 E_{\rm HFS}$,
$g_I$ and $g_J$ are the gyromagnetic ratios of the electron and nucleus, $\mu_B$ is the Bohr magneton, 
$E_{\rm HFS}$ is the hyperfine energy splitting at zero magnetic field \cite{Matuszewski_AF,Matuszewski_PS}.
Finally, the spin-dependent Hamiltonian (\ref{EA}) becomes
\begin{equation} 
H_{\rm A} = \int d x \, \left[ AB^2(\rho_+ + \rho_-) +\frac{c_2}{2} :{\bf F}^2: \right].
\end{equation}

\subsection{Ground states of the uniform system}
\label{sec:gsus}
The determination of ground states under the constraint of fixed magnetization is an interesting problem by itself and has been investigated by several authors, e.g. in~\cite{Zhang2003}.
In the following we briefly recall the results focusing on the system size much larger than the spin healing length $\xi_s=\hbar/\sqrt{2 m c_2 \rho}$.
In the case of a homogeneous system $V(x)=0$ one has to take into account the possibility of phase separation which occurs due
to the relation between the self- and cross-scattering terms in the Hamiltonian, as it has been observed experimentally~\cite{Ketterle_SpinDomains}. Let us define $b=B/B_0$ with $B_0=\sqrt{c_2 \rho/A}$.
Except for the special cases $M=0,\pm N$
\footnote{The ground state is the polar state ($\rho_0$) with all atoms in the $m_f=0$ component for $M=0$. 
Obviously, when $M=\pm N$, the system ground state is the ferromagnetic state ($\rho_\pm$) with all atoms in the $m_f=\pm1$ component.}, 
three types of ground states can exist divided by the two critical points at $b_1=M/(\sqrt{2} N)$ and $b_2=1/\sqrt{2}$. The ground state can be 
$(i)$ antiferromagnetic (2C) for $b < b_1$,
$(ii)$ phase-separated into two domains of the 2C and $\rho_0$ states for $b\in(b_1, b_2)$, or
$(iii)$ phase-separated into two domains of the $\rho_0$ and $\rho_+$ states for $b>b_2$~\cite{Matuszewski_PS}.
Moreover, the antiferromagnetic 2C state remains dynamically stable, up to a critical field $b_c>b_1$~\cite{ourPRB}. Consequently, the system driven adiabatically across the phase boundary $b_1$, from the 2C phase into the separated phase, remains in the initial 2C state up to $b_c^2=1-\sqrt{1-(M/N)^2}$, when the 2C state becomes dynamically unstable towards the phase separation.

\subsection{Ground states of the trapped system}
\begin{figure*}[]
\centerline{\includegraphics[width=0.25\textwidth,angle=-90]{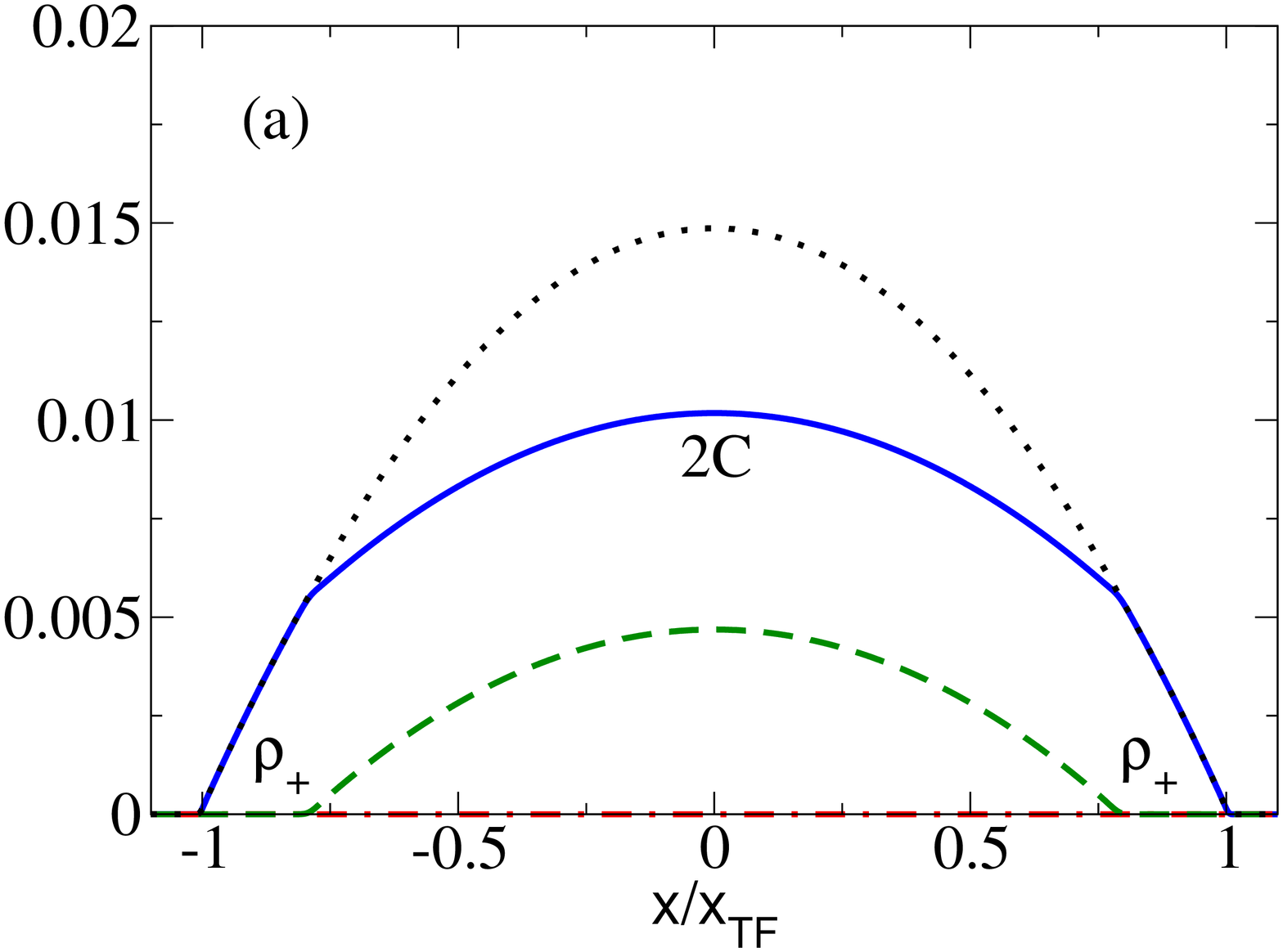}
\includegraphics[width=0.25\textwidth,angle=-90]{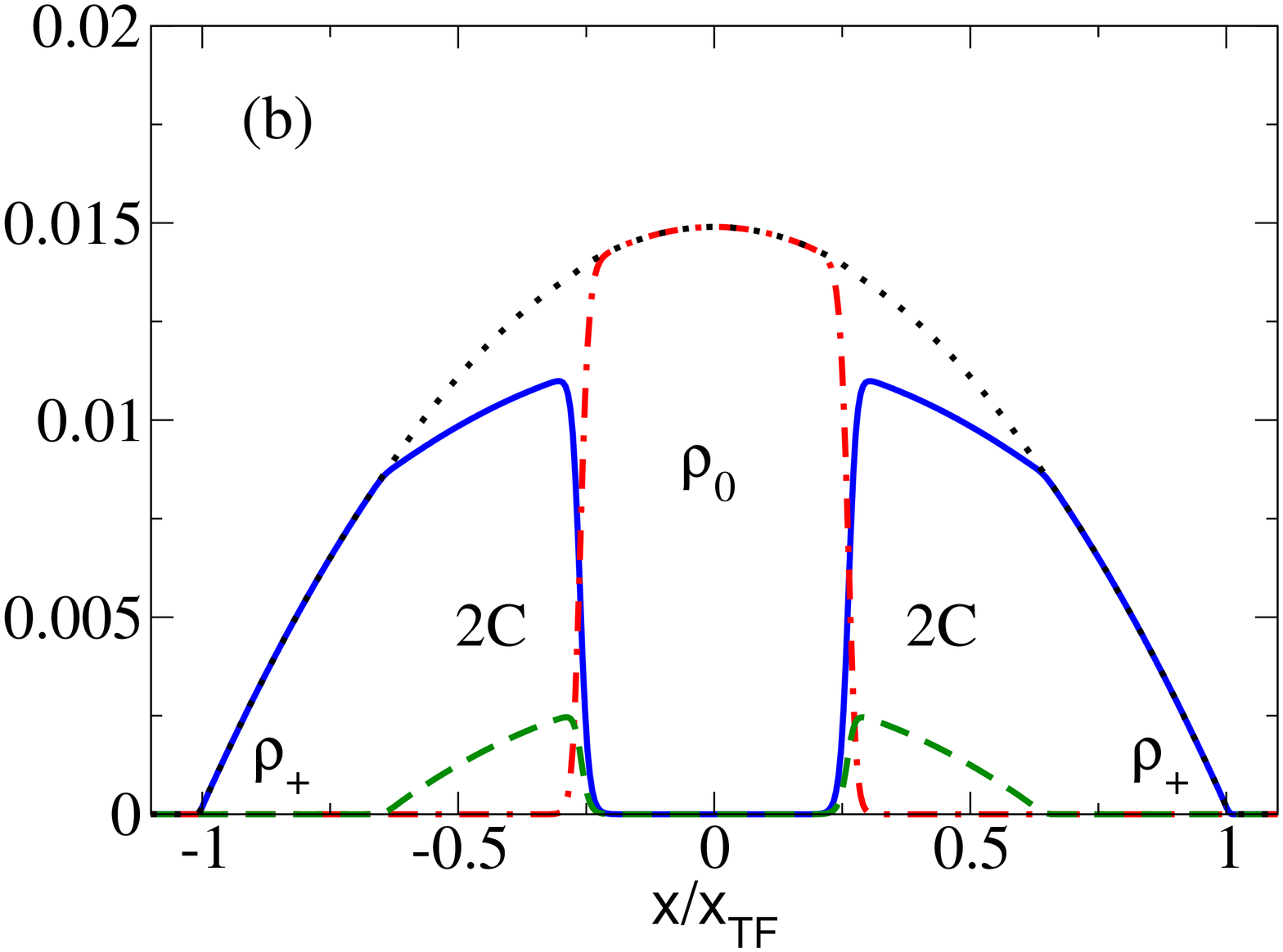}
\includegraphics[width=0.25\textwidth,angle=-90]{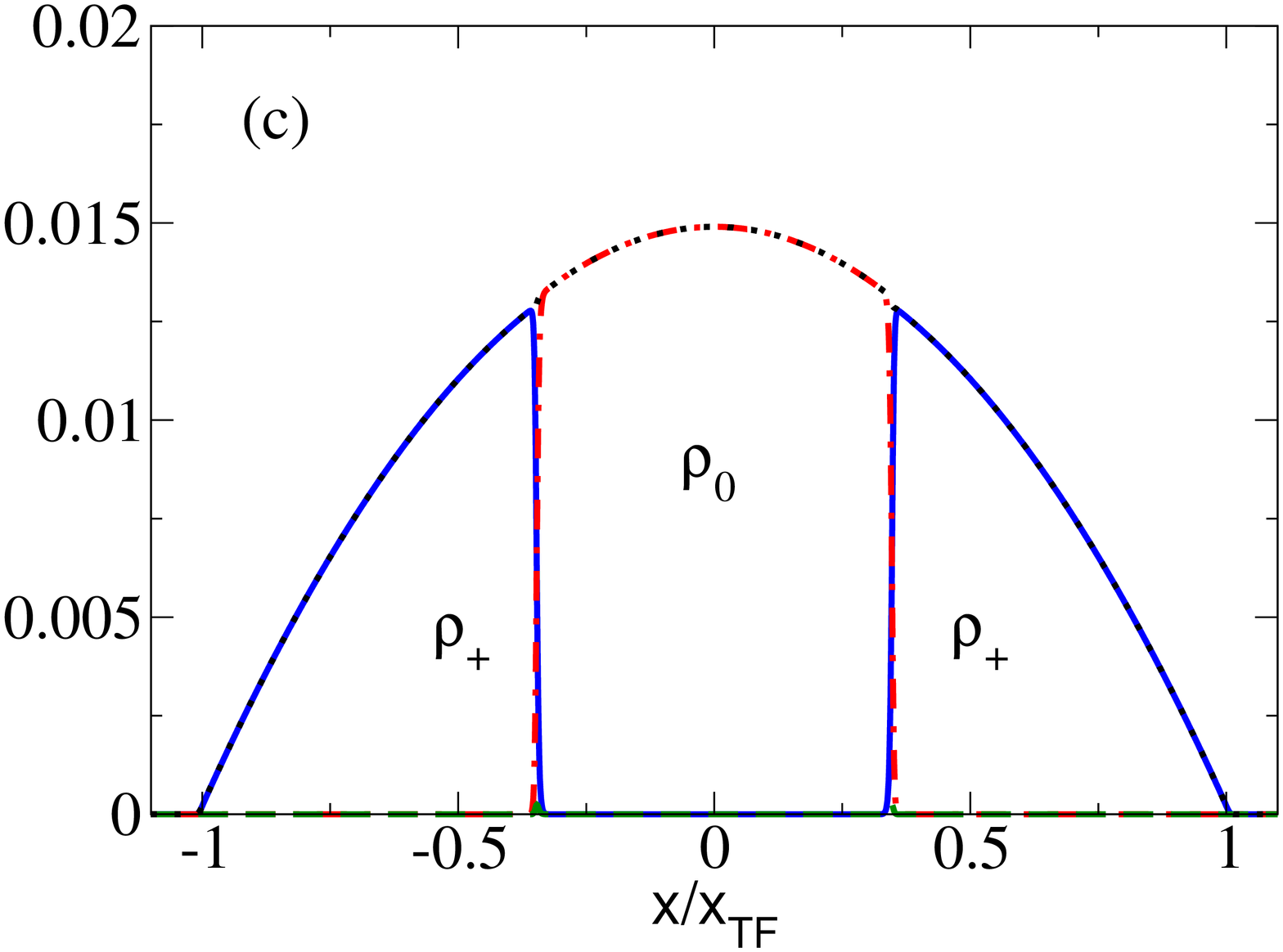}}
\caption{Density profiles of the $m_F=1$ (solid blue line), $m_F=0$ (dot-dashed red line) and $m_F=-1$ (dashed green line) components, and the total density (dotted black line). 
In the (a) ${\rm 2C}+\rho_+$ state for $b=0$, (b) $\rho_+ + {\rm 2C} + \rho_0$ state for $b=0.2$ and (c) $\rho_+ + \rho_0$ state for $b=0.5$, 
all for $N=2\times 10^6$, $\omega=2\pi\times 40$Hz, $\omega_\perp=2\pi\times1000$Hz and the magnetization $M=N/2$.}
\label{fig:fig1}
\end{figure*}

A more complicated situation occurs when the trap $V(x)=m \omega^2 x^2/2$ is present. 
The structure of ground states can be found analytically in the Thomas-Fermi approximation (TF)~\cite{Matuszewski_GroundStates, Gautam2015}, or numerically according to~\cite{Bao1,Bao2}.
We recall it in a regime of parameters such that the spin healing length $\xi_{\rm s}=\hbar/\sqrt{2mc_2\rho}$ is much smaller than the size of an atomic cloud determined by the TF radius. 
In the TF approximation, and under the assumption $c_0  \gg c_2$, the profile of the total density is independent of the magnetic field, 
\begin{equation}
\rho(x)=\rho(0)\left(1 - \frac{x^2}{x_{\rm TF}^2}\right)\, \,\,\, {\rm for} \, |x| \le x_{\rm TF},
\end{equation}
where $\rho(0)=m\omega^2 x_{\rm TF}^2/(2 c_0)$ and $x_{\rm TF}^3=3c_0N/(2 m \omega^2)$ is the Thomas-Fermi radius. 
However, density profiles of particular spin components depend both on the magnetization and the magnetic field, see Appendix \ref{app:TF} for explicit formulas.

The ground state can be $(i)$ the ${\rm 2C}+\rho_+$ state for $b < b_1$, shown in Fig.~\ref{fig:fig1}a,
$(ii)$ separated into the ${\rm 2C}+\rho_+$ and $\rho_0$ phases for $b_1 < b < b_2$, as shown in Fig.~\ref{fig:fig1}b, and 
$(iii)$ separated into the $\rho_+$ and $\rho_0$ phases for $b>b_2$, presented in Fig.~\ref{fig:fig1}c,
where we kept the notation introduced in~\cite{Matuszewski_GroundStates}. 
The presence of these states results from the interplay between phase separation and potential separation as shown in~\cite{Timmermans_PhaseSeparation}, and are absent for $\xi_{\rm s}$ larger than the TF radius. Spatial dependence of the two critical points can be derived in the local density approximation (LDA), as explained in Appendix \ref{app:B1B2}, and are
\begin{equation}
b_1(x) ^2 = \frac{1}{2}\left\{
\begin{array}{ccc}
\frac{(1-\frac{x_1^2}{x_{\rm TF}^2})^2}{1-\frac{x^2}{x_{\rm TF}^2}} &, \,\,\,\,& 0 \le |x|\le x_1, \\
1-\frac{x^2}{x_{\rm TF}^2}, & \,\,\,\,& x_1\le |x|\le x_{\rm TF} ,
\end{array}
\right.
\end{equation}
where $x_1^3=(1-M/N)x_{\rm TF}^3$, and
\begin{equation}
b_2(x) ^2 = \left\{
\begin{array}{ccc}
0, & \,\,\,\,& 0\le |x| \le x_2, \\
1-\frac{x^2}{x_{\rm TF}^2}, & \,\,\,\,& x_2\le |x|\le x_{\rm TF},
\end{array}
\right.
\end{equation}
with $x_2= x_{\rm TF} r_2$ and $r_2$ being a real, positive and smaller than one solution of the equation $r_2^3-3r_2+2(1-M/N)=0$.
In addition, the antiferromagnetic phase remains dynamically stable up to the critical field $b_c(x)$. Stability analysis of the initial 2C+$\rho_+$ state, which is based on the Bogoliubov transformation for the uniform system~\cite{ourPRB} treated with the LDA, gives 
\begin{equation}\label{eq:Bc}
b_c(x) ^2 =  \frac{\rho(x)}{\rho(0)}\left(  1- \sqrt{1 - \frac{m_{\rm 2C+\rho_+}(x)^2}{\rho(x)^2}} \right).
\end{equation}
The local magnetization of the 2C+$\rho_+$ state is
\begin{equation}\label{eq:mag}
m_{\rm 2C+\rho_+}(x) = \rho(0)\left\{
\begin{array}{ccc}
1- x_0^2, & \,\,\,\,& 0\le|x|\le x_1, \\
1-\frac{x^2}{x_{\rm TF}^2}, & \,\,\,\,&  x_1\le |x|\le x_{\rm TF}.
\end{array}
\right.
\end{equation}
It can be shown than $b_c(x)>b_1(x)$ for any $x$. Similarly as in the homogeneous system, there exists bistability of the $2C+\rho_+$ and $\rho_+ + {\rm 2C} + \rho_0$ states in the range $b\in\left[b_1, b_c\right]$. 

\section{Numerical experiment}
\begin{figure*}
\begin{picture}(0,100)
\put(-255,100){\includegraphics[width=0.2\textwidth,angle=-90]{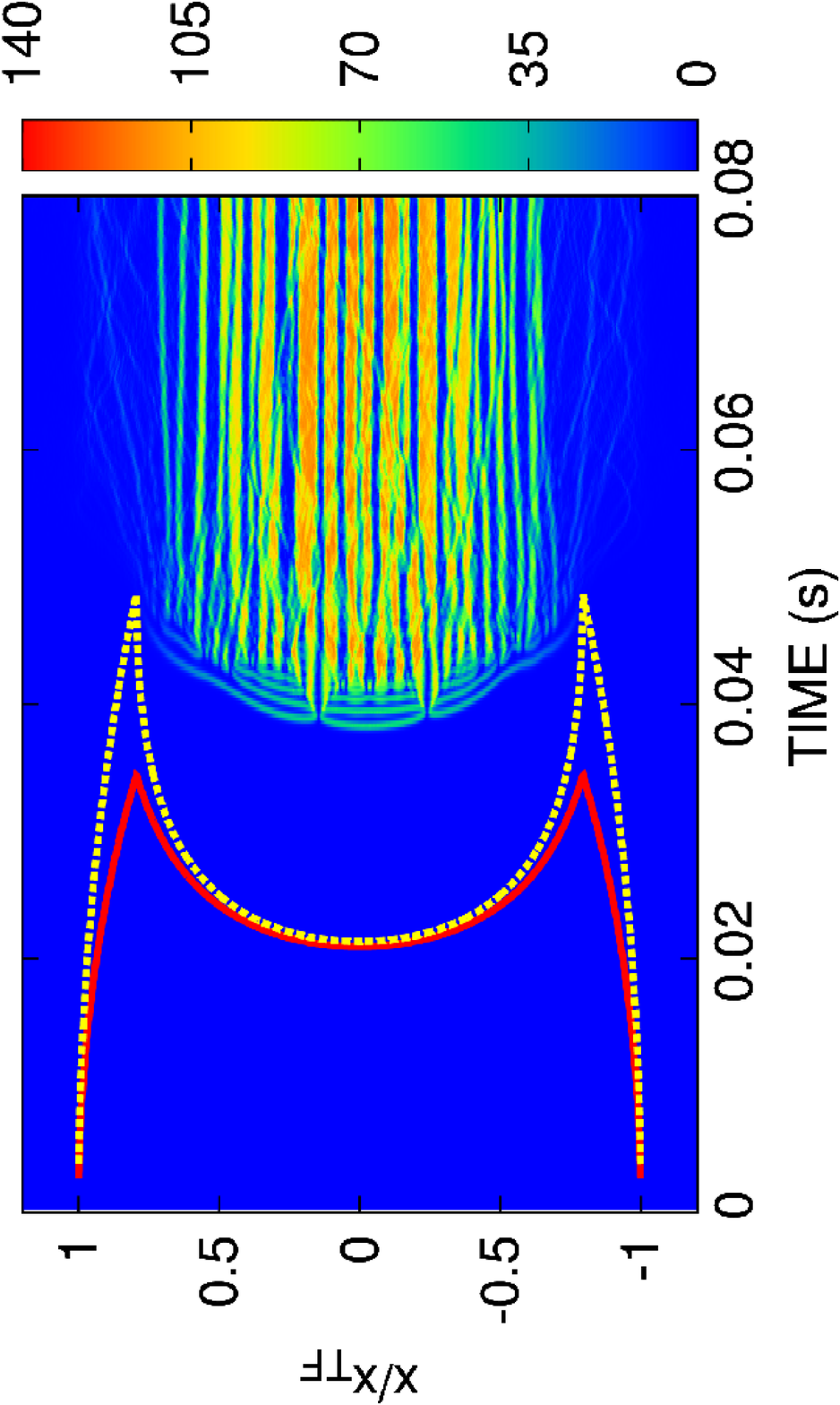}}
\put(-86,100){\includegraphics[width=0.2\textwidth,angle=-90]{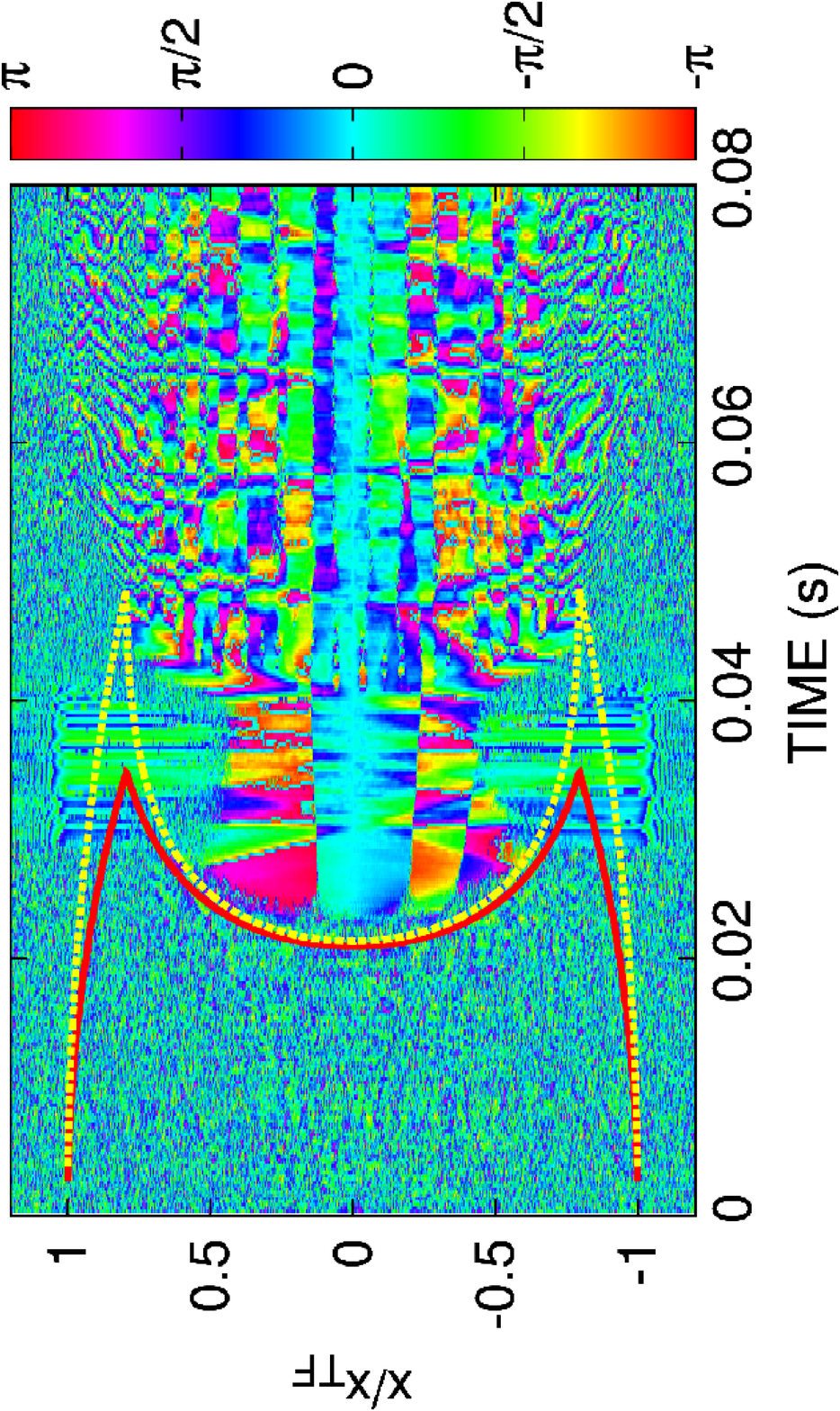}}
\put(85,100){\includegraphics[width=0.2\textwidth,angle=-90]{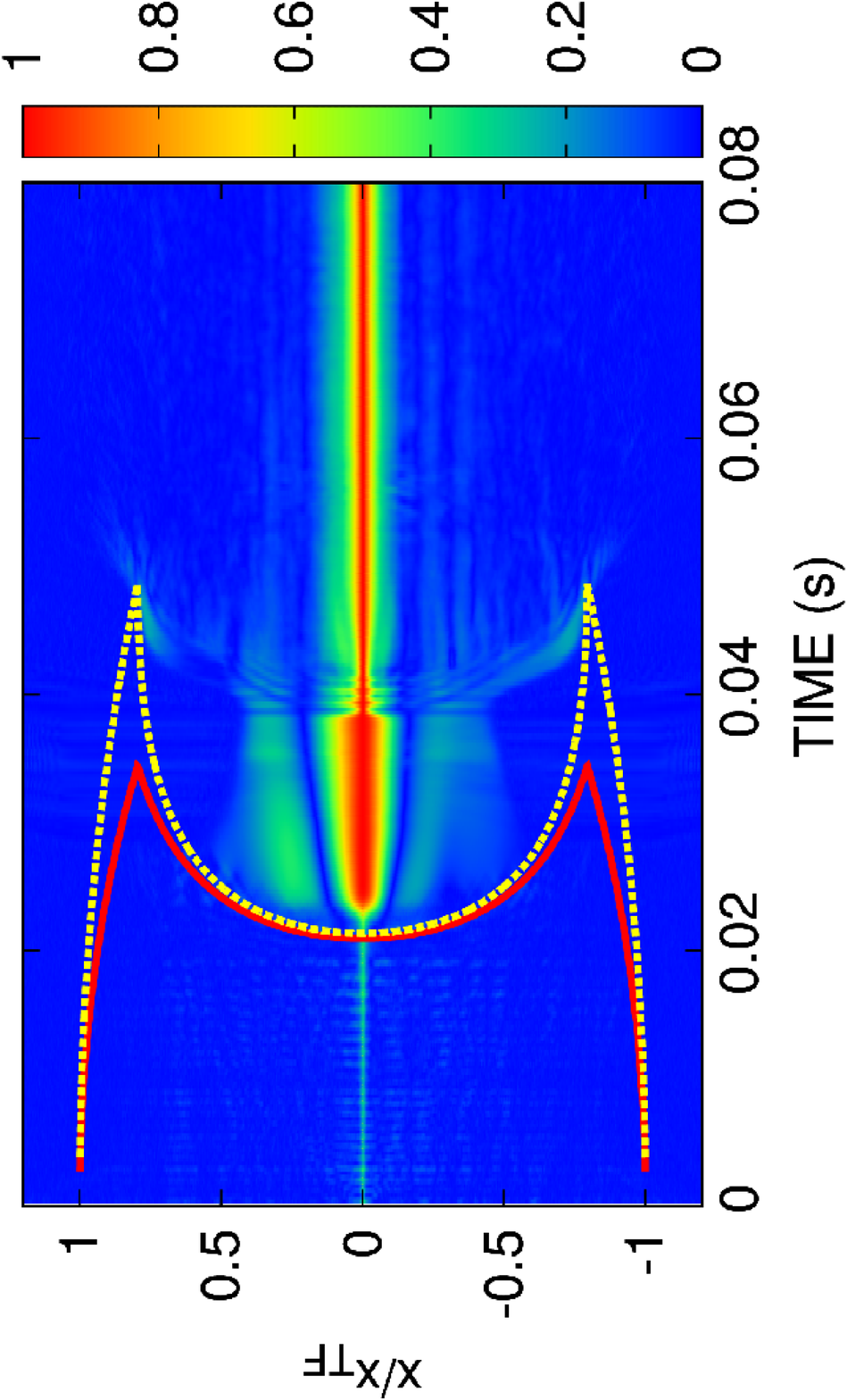}}
\put(-170,105){{\scriptsize{(a)}}}
\put(-5,105){{\scriptsize{(b)}}}
\put(165,105){{\scriptsize{(c)}}}
\end{picture}
\caption{The time evolution of the modulus square (a) and the phase (b) of $\psi_0(x,t)$ for a single realization of Wigner noise, as well as the correlation function of the first kind $g^{(1)}(x, t)$ given by Eq.~(\ref{eq:g1}) and averaged over $10^3$ realizations of Wigner noise (c). Lines show trajectories of the front $t_f(x)=\tau_Q b_f(x)$ with $f=1$ (solid red line), and $f=c$ (dashed white line). The other parameters are the same as in Fig.~\ref{fig:fig1}.}
\label{fig:fig2}
\end{figure*}

In order to investigate the effect of an external trapping potential on an outcome of the quantum phase transition in our system we performed numerical simulations by using the truncated Wigner method~\cite{WignerRef}. The initial state for the evolution is prepared using the numerical method proposed by Bao et al.~\cite{Bao1,Bao2} with additional stochastic noise added to mimic quantum fluctuations. In this way an ensemble of initial stochastic fields $\psi^T=(\psi_+,\psi_0,\psi_-)$ is prepared. The dynamics of every representative is governed by the coupled Gross-Pitaevskii (GP) equations
\begin{eqnarray} \label{Wigner}
i\hbar\frac{\partial \psi_0}{\partial t}  &=& \left(-\frac{\hbar^2 \nabla^2}{2m} + \frac{1}{2}m \omega^2 x^2 + c_0\rho\right)\psi_0 + \nonumber\\
      &&+c_2\left[(\rho_1+\rho_{-1})\psi_0+2\psi_0^*\psi_1\psi_{-1}\right],\nonumber\\
i\hbar\frac{\partial \psi_1}{\partial t}   &=& \left(-\frac{\hbar^2 \nabla^2}{2m} +\frac{1}{2}m \omega^2 x^2 + c_0\rho + A B^2\right)\psi_1 + \nonumber\\
      &&+c_2\left[(\rho_1-\rho_{-1})\psi_1+\rho_0\psi_1+\psi_{-1}^*\psi_0^2\right],\\
i\hbar\frac{\partial \psi_{-1}}{\partial t}   &=& \left(-\frac{\hbar^2 \nabla^2}{2m} + \frac{1}{2}m \omega^2 x^2 +c_0\rho + A B^2\right)\psi_{-1} + \nonumber\\
      &&+c_2\left[(\rho_{-1}-\rho_1)\psi_{-1}+\rho_0\psi_{-1}+\psi_1^*\psi_0^2\right],\nonumber
\end{eqnarray}
which follow from the spinor Hamiltonian.
The initial state for the evolution is the antiferromagnetic ground state 2C+$\rho_+$ for $b=0$. Next, the magnetic field is increased linearly in time
\begin{equation}
b(t)=\frac{t}{\tau_Q},
\end{equation}
were the final magnetic field $b(\tau_Q)$ is larger than $b_c(x_1)$. Above $b_c$ the system is expected to undergo the spatial symmetry breaking phase transition. 
The example of time evolution is given in Fig.~\ref{fig:fig2} for $\tau_Q=80$ms, where the modulus square and the phase of $\psi_0(x,t)$ are plotted in addition to the first order correlation function
\begin{equation}\label{eq:g1}
g^{(1)}(x, t)=\frac{\langle \psi_0^*(x,t) \psi_0(0,t) \rangle}{\sqrt{\langle |\psi_0(x,t)|^2 \rangle \langle |\psi_0(0,t)|^2 \rangle}},
\end{equation}
were averages are taken over stochastic realizations of the Wigner noise.

Closer investigation of the results allows us to make several interesting observations about the outcome of the non-adiabatic and inhomogeneous phase transition. Spin domains appear, and further post-selection of them is clearly visible, as illustrated by dthe ensity of the $m_f=0$ component in Fig.~\ref{fig:fig2}a. At the same time, the phase of $\psi_0$ experiences sudden jumps at positions of the domain walls, and phase domains that appear are of size comparable to the separation between neighboring walls, see Fig.~\ref{fig:fig2}b. This demonstrates weak coherence between created spin domains, and the final configuration of spin domains can be seen as a set of quasi-condensates. In addition, spin domains nucleate and remain in the region limited by the $\rho_0$ phase size, from $-x_1$ to $x_1$. The presence of the bound is an advantage in the domain number counting, since it allows one to avoid counting at the boundary of the system where the domain may be comparable to noise. In parallel to the domain formation, the process of transport of local magnetization from the center to boundaries of the system is pronounced, see Fig.~\ref{fig:fig4}. The origin of the effect lies in the sign of $c_2$ in the interaction energy (\ref{EA}) which favors locations of domains having zero local magnetization in the trap center where the density is largest~\cite{Matuszewski_GroundStates}. The opposite direction of the local magnetization transport may be expected for ferromagnetic spinor condensates for which the sign of $c_2$ is negative. Another characteristic feature is the emergence of spin domains from spin waves which can be observed in the density plot of the $m_f=0$ component in Fig.~\ref{fig:fig2}a. The presence of spin waves results from the coherent spin mixing dynamics~\cite{Zhang2005}, so it is specific for spinor condensates. Initially, a very tiny order parameter appears in the $m_f=0$ component above $b_c$ through the coherent process $\rho_+ + \rho_-\to \rho_0$. During this very short period of time the phase of $\psi_1\psi_{-1}$ is imprinted on $\psi_0^2$. The sudden increase of the phase coherence of the $m_f=0$ component can be observed just above $t_c$ and before domains nucleation, as illustrated by the phase of the $m_f=0$ component in Fig.~\ref{fig:fig2}b and the correlation function in Fig.~\ref{fig:fig2}c. When domains are formed range of the correlation function changes and corresponds to the final average size of domains. 

According to the KZ mechanism, the system ends up in a state with multiple spin domains. The concept of the mechanism relies on the fact that during the non-adiabatic quench the system does not follow the ground state exactly in a vicinity of the critical point. This is due to the divergence of the relaxation time. In the uniform system the quantum phase transition from an antiferromagnetic to phase separated state exhibits two scaling laws~\cite{ourPRL}. The KZ theory results in scaling laws $N_d\sim\tau_Q^{-1/3}$, coming from critical exponents $\nu=1/2$ and $z=1$, for density of spin domains seeds that are formed just after crossing the critical point. Further on, the post-selection process forced by the conservation of magnetization takes place, leading to the second scaling law with a different exponent $N_d\sim\tau_Q^{-2/3}$. 

The inhomogeneity, arising as a result of the external trapping potential, brings new ingredients. Due to the spatial dependence of $b_c(x)$ different parts of the system undergo phase transition at different times as the magnetic field grows up from zero. As the result, the relaxation time and correlation length acquire local dependence. It is widely understood that the domain formation is governed by causality. When the front of the transition moves faster than the characteristic velocity of the perturbation, domains nucleate. Otherwise the choice of the order parameter in the broken symmetry phase is done homogeneously along the system. The effect of the moving front changes the qualitative result of the KZ theory and scaling exponents as well \cite{Zurek2009, Dziarmaga2010, Sabbatini2012, delCampo2013, Saito2013, delCampo2014}. 

\begin{figure}
\includegraphics[width=0.45\textwidth]{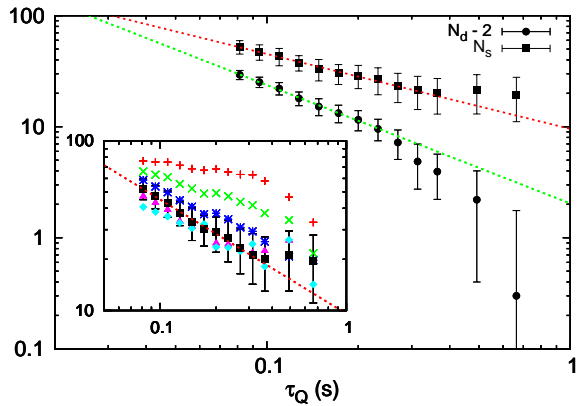}
\caption{Scaling of the number of domain seeds $N_s$ (points) just after crossing the critical point $b_1$, and of the number of domains $N_d$ (boxes) in final stable configurations; 
scaling $N_s\propto \tau_Q^{-2/3}$ is denoted by the red dashed line while $N_d\propto \tau_Q^{-1}$ by the green dashed line. The instet shows the number of domain seeds obtained for different values of the counting threshold, $\alpha=0.03,\,0.04,\,0.05,\,0.06,\,0.07,\,0.25$ from top to bottom, demonstrating quite a wide range of the scaling validity. Here the average is taken over $10^2$ realizations of the Wigner noise, the other parameters are the same as in Fig.~\ref{fig:fig1}.}
\label{fig:fig3}
\end{figure}

\begin{figure*}[]
\begin{picture}(0,100)
\put(-255,0){\includegraphics[width=0.322\textwidth]{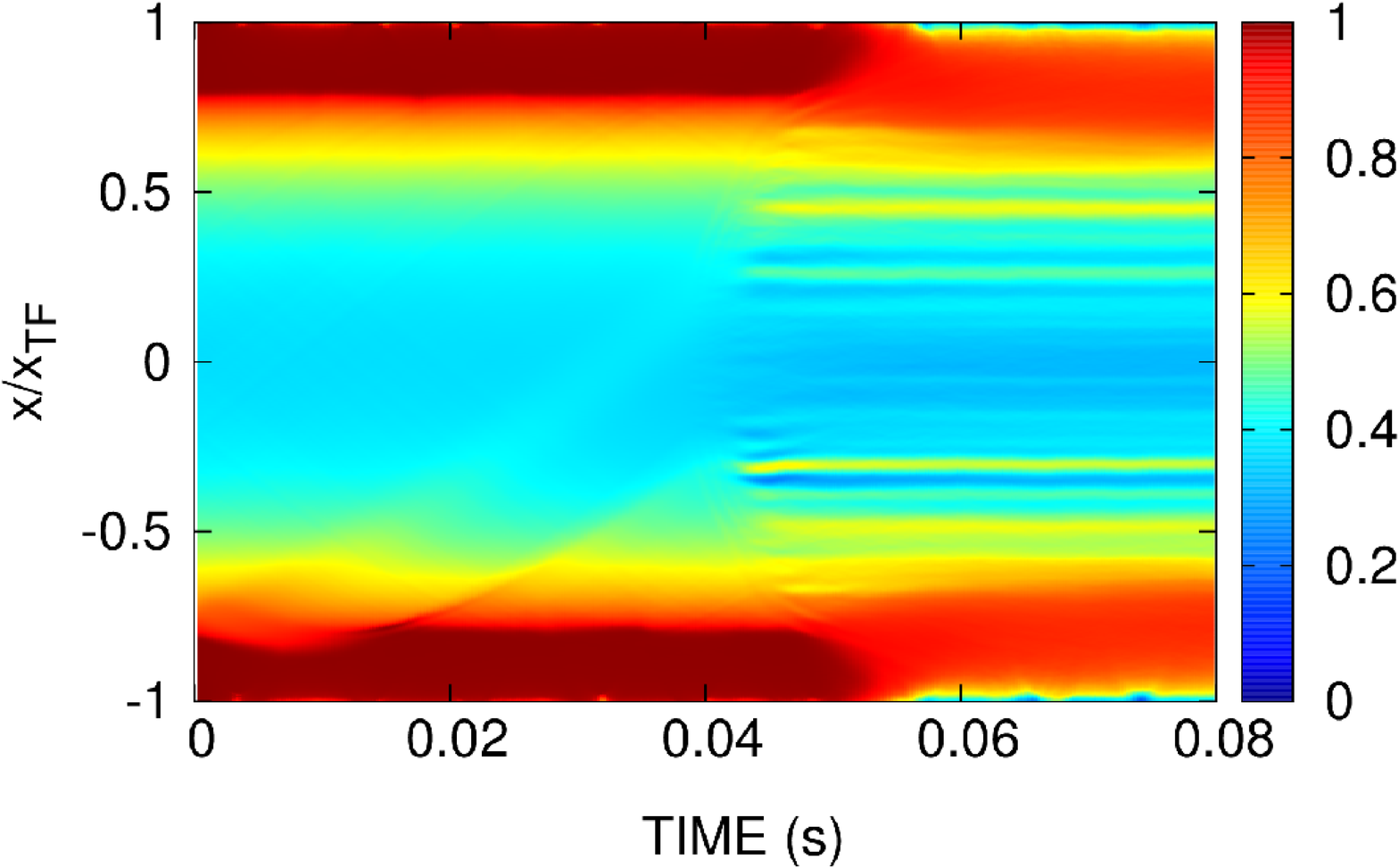}}
\put(-86,100){\includegraphics[width=0.2\textwidth,angle=-90]{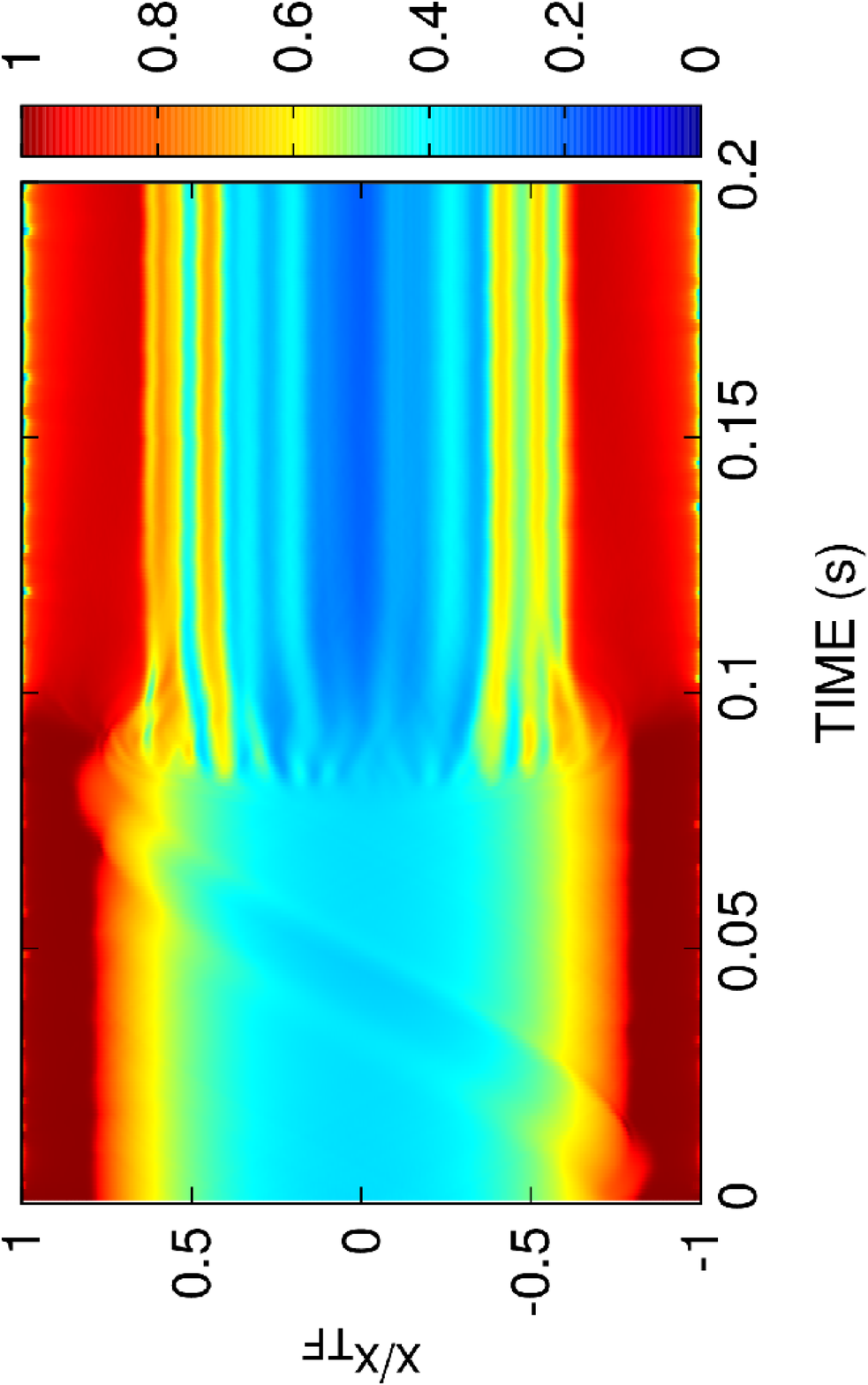}}
\put(85,100){\includegraphics[width=0.2\textwidth,angle=-90]{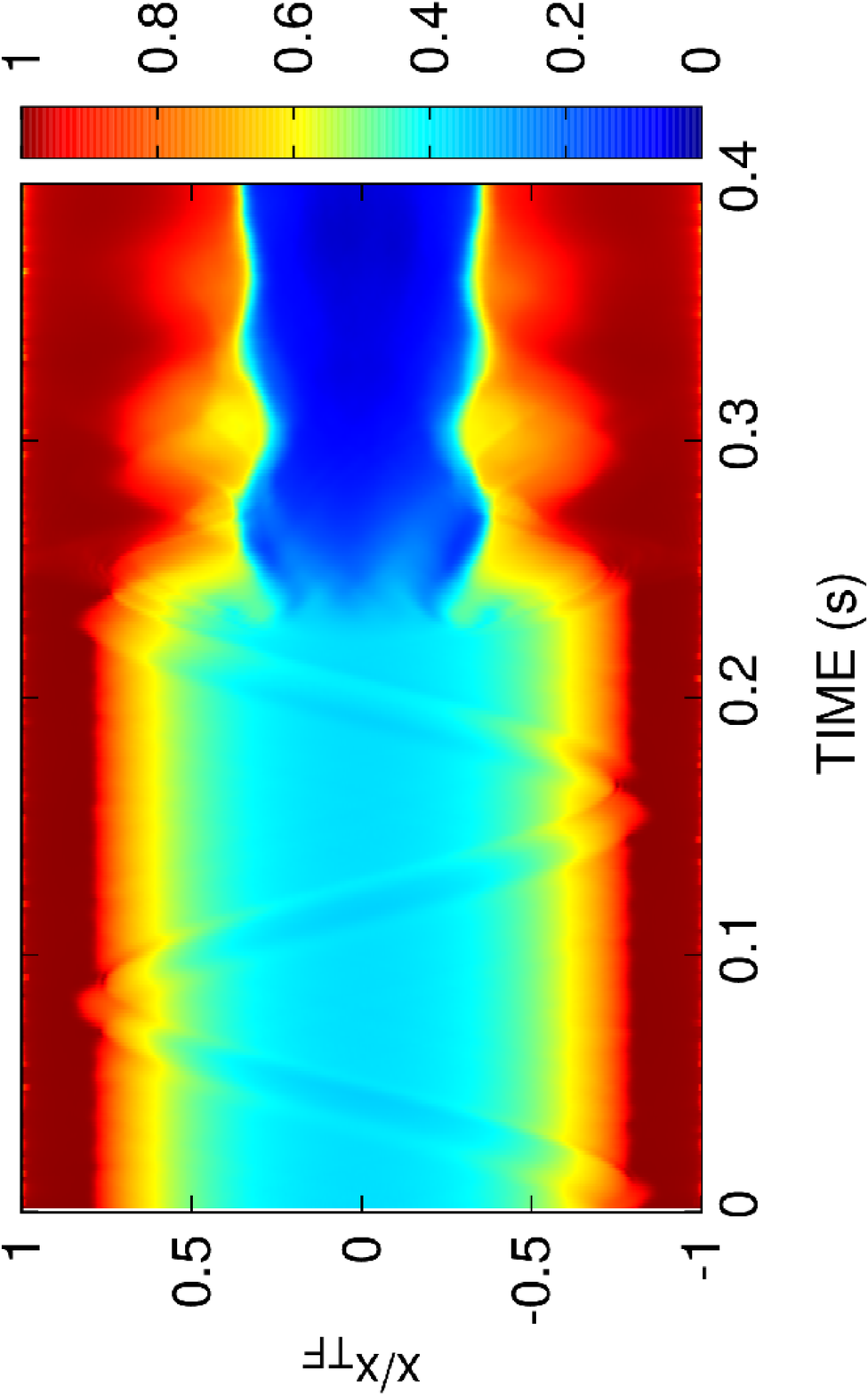}}
\put(-175,105){{\scriptsize{(a)}}}
\put(-5,105){{\scriptsize{(b)}}}
\put(165,105){{\scriptsize{(c)}}}
\end{picture}
\caption{Density of the local magnetization divided by the local density $\langle m(x,t)\rangle/\rho(x)$ averaged over $3\times 10^3$ realizations of the Wigner noise for $\tau_Q=80$ms (a), $\tau_Q=200$ms (b) and $\tau_Q=700$ms (c), other parameters are the same as in Fig.~\ref{fig:fig1}. Pushing the local magnetization out of the trap center after domain seeds formation is clearly visible.}
\label{fig:fig4}
\end{figure*}

Numerical results for scaling of the domains number\footnote{To determine the number of domains $N_d$ or domain seeds $N_s$ we count the number of zero crossings of  the function $f(x)=\rho_0(x)-\alpha \rho(x)$, where the best choice is $\alpha=0.5$ for domains and $\alpha=0.06$ for domain seeds, at the time instant when $N_s$ or  $N_d$ is the largest. We checked that this method is accurate and weakly dependent on the choice of $\alpha$.} are presented in Fig.~\ref{fig:fig3}. The result shows scaling of the domain seeds number just after crossing the critical point $b_c$ to be $N_s\propto \tau_Q^{-2/3}$, and scaling of the domains number in final stable configurations to be $N_d\propto \tau_Q^{-1}$. We emphasize that these scaling laws cannot be explained by the causality effect alone. In the next section we present an analytical treatment for the two scaling laws derivation based on a slow diffusive-drift relaxation process of the local magnetization transport.

\section{Two scaling laws}

We consider the phase transition from an antiferromagnetic ground state, it is the 2C phase for the uniform system, to a phase separated state by linearly increasing the magnetic field $b=t/\tau_Q$. The distance from the critical point is measured by a dimensionless parameter
\begin{equation}\label{eq:epsilon}
\epsilon(t)=b - b_c,
\end{equation}
and is a linear function of time $\epsilon\sim t/\tau_Q$ (here we choose $t=0$ at the first critical point). 

Before we proceed into details let us briefly remind our results for the uniform system~\cite{ourPRB}. 

We analyzed the initial antiferromagnetic ground state in the Bogoliubov approximation. The Bogoliubov spectrum is composed of three branches wherein one gapped is
\begin{equation}
\frac{\omega_k^2}{(c_2 \rho(0))^2} = \left(\xi_s^2 k^2 +\tilde{\rho} - b^2 \right)^2 - \left( \tilde{\rho} - b^2_c\right)^2,
\end{equation}
where $\tilde{\rho}=\rho/\rho(0)$. The critical magnetic field $b_c$ is obtained from $\omega_0^2=0$. Here we have written the gapped spectrum in the form convenient for our further LDA analysis. Notice that $\rho(0)=N/L$, $\tilde{\rho}=1$ for the homogeneous system of size $L$. While increasing the magnetic field from zero, the 2C phase remains dynamically stable up to $b_c$.

The reaction time of the system $\tau_r\sim \omega_0^{-1}$ is the shortest time scale on which the ground state of the system can adjust adiabatically to varying $b$. The energy gap vanishes as $\omega_0\sim |\epsilon|^{z\nu}$, so the evolution across the critical point cannot be adiabatic. 
The KZ theory \cite{Zurek} is based on an approximation in which time evolution near the critical point is divided into three stages. The border between particular stages is the time instant $\hat{t}$ at which the reaction time of the system is comparable to the transition time $\tau_t=|\epsilon/\dot{\epsilon}|$. The equality $\hat{\tau}_r=\hat{\tau}_t$ defines $\hat{t}$, which for our system is $\hat{t}\sim\tau_Q^{1/3}$. So, at $t=-\hat{t}$ the state of the system is assumed to be an adiabatic ground state with a correlation length $\hat{\xi}$. This state freezes-out at $-\hat{t}$, and does not change till $\hat{t}$,
\footnote{This point is violated a bit in our case, since Bogoliubov modes become unstable for $t>t_c$ and corresponding fluctuations blow up.}. 
At $t=\hat{t}$ the frozen state is no longer the ground state but an excited state with the correlation length $\hat{\xi}$, becoming the initial state for further adiabatic evolution. The average number of domain seeds is determined by the correlation length at the freeze-out time $N_s=L/\hat{\xi}$ which is set by the maximal unstable momentum mode at $\hat{t}$, $\hat{\xi}=1/\hat{k}$, and from $\omega_{\hat{k}}(b=\hat{b})=0$ one has
\begin{equation}
\xi_s\hat{k}\simeq \sqrt{b_c \hat{\epsilon}}.
\end{equation}
The number of spin domain seeds scales as $N_s\sim \tau_Q^{-1/3}$ for our system. 

The standard KZ scenario is strongly modified by the post-selection process forced by the conserved magnetization in our system, leading to the second scaling law (see \cite{ourPRB} for a detailed explanation of the process). The derivation of the second scaling exponent required an observation that the number of domains in a stable configuration is determined by the fraction $x_0$ of the system occupied by the $\rho_0$ phase divided by the healing length at the freeze-out time, $N_d=\hat{x}_0/\xi_{2C + \rho_0}$ with $\xi_{2C + \rho_0}=\xi_s/b^2$. This healing length is finite near the phase transition. In \cite{ourPRL} we have shown that $x_0=1-b_1/b$, which indicates that $\hat{x}_0\sim\hat{\epsilon}$ for weak magnetization as $b_c\simeq b_1$. The second scaling law is then $N_d\sim \tau_Q^{-2/3}$.

A standard treatment of the analysis of scaling laws in a trapped system is based on the local density approximation. From now on we consider the spatially dependent critical magnetic field $b_c(x)$, magnetization $m(x)=\rho_+(x)-\rho_-(x)$ and distance from the critical point $\epsilon(x, t)=b(t) - b_c(x)$, for $x\le x_1$. The number of spin domain seeds as well as domains in a stable configuration can be estimated as follows (up to some numerical factors $f_i$):
\begin{equation}\label{eq:integrals}
N_s\equiv \int_{-\hat{x}_{\rm max}}^{\hat{x}_{\rm max}} \frac{dx}{f_s \hat{\xi}},~~~ N_d\equiv \int_{-\hat{x}_{\rm max}}^{\hat{x}_{\rm max}} \frac{\hat{x}_0 (x) dx}{f_d \xi_s},
\end{equation}
where the integration runs over the reduced length $\hat{x}_{\rm max}$ which we will explain below. 
If one neglects $x$-dependence of $\hat{\xi}$ and $\hat{x}_0$, then the expression for the number of defects simplifies even more, $N_s\simeq 2\hat{x}_{\rm max}/(f_s \hat{\xi})\propto \hat{x}_{\rm max} \tau_Q^{-1/3}$ and $N_d\simeq 2\hat{x}_0/(f_d \xi_s)\propto \hat{x}_{\rm max} \tau_Q^{-2/3}$. 
The $\tau_Q$ dependence of $\hat{x}_{\rm max}$ remains to be established in order to determine the KZ scaling of the defects density. It is easy to notice that $\hat{x}_{\rm max}\sim \tau_Q^{-1/3}$ matches results of our numerical experiment, as shown in Fig.~\ref{fig:fig3}. 

In the case of our system there are two processes which lead into limited by $\hat{x}_{\rm max}$ range of the defect formation, namely the causality and transfer of the local magnetization from the trap center to the system boundaries. We emphasize that the latter effect establishes the desired scaling of $\hat{x}_{\rm max}$.

\subsection{Causality}

The causality sets $\hat{x}_{\rm max}$ at the freeze-out time to the space region where the front of the transition $v_F$ moves faster than any perturbation $v$. The characteristic velocity of perturbation can be upper bounded by the ratio of the correlation length over the relaxation time, and at the freeze-out time it is $\hat{v}=\hat{\xi}/\hat{\tau}_r$~\cite{delCampo2011}. It does not depend on $\tau_Q$, whereas it still has some space dependence which we treated within the LDA. The speed of the front is $v_F=|d t_c/dx|^{-1}$, where $t_c=\tau_Q b_c$. The scaling of $\hat{x}_{\rm max}$ is set by the equality $\hat{v}=\hat{v}_F$ which we write symbolically as $\tau_Q^{-1}=f(\hat{x}_{\rm max})$, where $f(\hat{x}_{\rm max})$ is a know function independent of $\tau_Q$. The expansion of $f(\hat{x}_{\rm max})$ in the Taylor series up to the leading term in $\hat{x}_{\rm max}$ gives the scaling $\hat{x}_{\rm max}\propto \tau_Q^{-1}$ which is not the observed one.

\subsection{Transport of the local magnetization}

Transport of the local magnetization during the formation of domains is a quite complicated process, but we can estimate the modification of scaling laws if we make a few assumptions. We focus on the area close to the center of the trap where most of domains are created, then $x/x_{\rm TF}$ is a small parameter of our theory; and on  weak magnetization, then the two critical fronts are indistinguishable $t_c(x)\approx t_1(x)$ and also $\hat{t}(x)\approx t_1(x)$.
The critical point is first crossed in the center of the trap, where the 2C phase is initially present, see Fig.~\ref{fig:fig1}(a). This results in the production of $\rho_0$ atoms from the 2C phase via the process $\rho_- + \rho_+ \rightarrow 2\rho_0$ through contact interactions. The $\rho_0$ and 2C phases are repelled apart when the system undergoes phase separation. Energetic considerations point out that the preferred state is the one where the $\rho_0$ phase is situated in the center of the trap, while the magnetized 2C phase resides away from the trap center~\cite{Matuszewski_GroundStates}. This will lead to the transport of local magnetization across the system. However, the process of local domain seeds formation is, for realistic quench times, faster than the transport of local magnetization. For this reason we will consider the transport of local magnetization as a slow process, which nevertheless can lead to a modification of the scaling exponents. To illustrate the effect of the local magnetization transport, in Fig.\ref{fig:fig4} we show examples of the time evolution of the averaged local magnetization. 

Now, we make the central assumption that the transport of local magnetization is a slow diffusive-drift relaxation process. The local magnetization at the point $x$, which we denote by $m_x(t)$, changes in time due to its transport from the trap center to $x$. We estimated its evolution from the drift-diffusion equation
\begin{equation}\label{eq:cde}
\frac{\partial m_x(t)}{\partial t} \sim -\nabla \cdot \left( m_x(t) u \right),
\end{equation}
where $u$ is the average velocity that the local magnetization moves with. Note, that the magnetization is $x$-independent for $b<b_1$ and $x<x_1$, see (\ref{eq:mag}). In the equation we neglected diffusion, sources or sinks and kept only the leading drift term. 
An approximate solution for very short times $\Delta t$ is
\begin{equation}\label{eq:dm}
m_x(\Delta t) - m_x(0) \sim -(\nabla \cdot u ) \Delta t.
\end{equation}
To describe the velocity of local magnetization we adopt and generalize the method of~\cite{Ketterle_TunnelingDomains} devised for the description of quantum tunneling across domains. The velocity of local magnetization is, in general, $u=\nu_u \Delta \mu$, where $\nu_u$ is the mobility which we will assume to be a constant in the lowest order approximation. The transfer of magnetization is enforced by a difference in chemical potentials of the growing neighbouring phases $\rho_0$ and 2C. In the pure $\rho_0$ phase, we denote the local chemical potential of the $m_f=0$ atoms (the cost of adding another $m_f=0$ particle to the $\rho_0$ phase) by $\mu_0$, and in the 2C phase by $\mu_{2C}$. At the critical point the chemical potential difference $\Delta \mu=\mu_{2C}- \mu_0$ is exactly zero and becomes positive in the phase-separated regime increasing to the first order in the small parameter $\Delta\mu(x,t)\sim \epsilon(x,t)$, so linearly with magnetic field according to (\ref{eq:epsilon}). 
Therefore, we approximate the velocity gradient as
\begin{equation}
\nabla \cdot u \sim \frac{\Delta u}{\Delta x} \sim \frac{\Delta\mu(x,t) - \Delta\mu(0,t)}{|x|} .
\end{equation}

The dynamics cease to be adiabatic at the freeze-out time $\hat{t}(x)$. The maximum distance at which domains are formed $\hat{x}_{\rm max}$, at the freeze-out time, is approximately the $x$ at which the change of local magnetization becomes comparable to the local density. No more domains can form beyond this point. Therefore, $m_{\hat{x}}(\Delta \hat{t}) - m_{\hat{x}}(0)\sim1$ at $\hat{x}=\hat{x}_{\rm max}$ defines scaling of ${\hat{x}_{\rm max}}$ with the quench time $\tau_Q$. We approximate $\Delta\hat{t}\simeq t_1(\hat{x})-t_1(0)\sim \hat{x}^2 \tau_Q$ and $\Delta \mu(\hat{x},\hat{t}) - \Delta \mu(0,\hat{t})\simeq b_1(0) - b_1(\hat{x})\sim \hat{x}^2$ according to our assumption of the two transition fronts to be comparable, $b_1\sim b_c$. Expanding the right hand side of (\ref{eq:dm}) at the freeze-out time up to the leading terms in $\hat{x}/x_{\rm TF}$ gives
\begin{equation}
m_{\hat{x}}(\Delta \hat{t}) - m_{\hat{x}}(0) \simeq \left| \hat{x} \right|^3 \tau_Q
\end{equation}
and the desired scaling with quench rate, $\hat{x}_{\rm max} \sim \tau_Q^{-1/3}$.

\section{Summary}
The KZ theory is a powerful tool that allows predicting the average size of domains forming topological defects resulting from a non-adiabatic phase transition without solving the full dynamical equations. However, the theory should be more developed in some specific cases when processes changing scaling exponents occur in the system. 

The antiferromagnetic spinor condensate turns out to be a very interesting case. Double universality in the dynamics takes place, and two scaling laws appear, not one as usually. It is the effect of the post-selection process forced by conserved magnetization which determines the density of the new phase and the density of spin domains in final stable configurations. The trapped system reveals additional modifications due to the causality and the transport of magnetization processes. The latter effect, characteristic for spinor condensates, imposes a stronger limit on the area in which spin domains can form. We consider both mechanisms, the post-selection process and the magnetization transfer across the system, to be general and effective whenever the standard KZ mechanism is not compatible with an additional conservation law.

It would be very instructive to examine experimentally modification of the KZ mechanism by conserved magnetization, while an experimental verification of the scaling exponents for the number of defects is still a challenge. Although, the recent experiment~\cite{Chapman2016} confirms the KZ theory for the scaling of the time instant $\hat{t}$, showing some minor modification of the scaling exponent resulting from atomic losses in the system. The KZ theory of spin-1 systems is quite well developed, however the change of scaling of the number of domains due to additional effects like particle losses, phase ordering kinetics or reduced dimensionality need to be further investigated, providing an interesting direction for future work.

\acknowledgments
The authors would like to acknowledge J. Dziarmaga for his initial contribution to the project. E.W. acknowledges discussion with B. Damski and O. Hul for a careful reading of the manuscript.
This work was supported by the National Science Center Grants DEC-2015/18/E/ST2/00760, DEC-2012/07/E/ST2/01389, DEC-2011/01/D/ST3/00482, and DEC-2015/17/D/ST2/03527.

\appendix

\section{Density profiles of particular phases in the TF approximation}
\label{app:TF}

Here and below we summarize simple expressions for ground state density profiles of particular components which result from the TF analysis. All of them were checked against numerical results by applying a method proposed by Bao et.al.~\cite{Bao1,Bao2}.

(i) In the 2C+$\rho_+$ state one has:
\begin{equation}
\frac{\rho_1(r)}{\rho(0)} = \left\{
\begin{array}{ccc}
1-\frac{1}{2}\left( r_1^2 + r^2\right), & \,\,\,\,& |r|\in(0, r_1), \\
1-r^2, & \,\,\,\,& |r|\in (r_1, 1) ,
\end{array}
\right.
\end{equation}
\begin{equation}
\frac{\rho_{-1}(r)}{\rho(0)} = \left\{
\begin{array}{ccc}
\frac{1}{2}\left( r_1^2 - r^2\right), & \,\,\,\,& |r|\in(0, r_1), \\
0, & \,\,\,\,& |r|\in (r_1, 1) ,
\end{array}
\right.
\end{equation}
and $\rho_0(r)=0$ for any $r$.
Here $\rho(0)=m\omega^2 x_{TF}^2/(2 c_0)$, $r=x/x_{TF}$, $x_{TF}^3=3c_0N/(2 m \omega^2)$ and $r_1^3=1-M/N$ (it was derived from the difference $N-M$).

(ii) the $\rho_+$+ 2C + $\rho_0$ state one has the following densities in particular components:
\begin{equation}
\frac{\rho_1(r)}{\rho(0)} = \left\{
\begin{array}{ccl}
0, & \,\,\,\,& |r|\in (0, u_1) , \\
1-\frac{1}{2}\left( u_0^2 + r^2\right), & \,\,\,\,& |r|\in(u_1, u_0), \\
1-r^2, & \,\,\,\,& |r|\in (u_0, 1) ,
\end{array}
\right.
\end{equation}
\begin{equation}
\frac{\rho_0(r)}{\rho(0)} = \left\{
\begin{array}{ccl}
1 - r^2, & \,\,\,\,& |r|\in(0, u_0), \\
0, & \,\,\,\,& |r|\in (u_0, 1) ,
\end{array}
\right.
\end{equation}
and
\begin{equation}
\frac{\rho_{-1}(r)}{\rho(0)} = \left\{
\begin{array}{ccl}
0, & \,\,\,\,& |r|\in (0, u_1) , \\
\frac{1}{2}\left( u_0^2 - r^2\right), & \,\,\,\,& |r|\in(u_1, u_0), \\
0, & \,\,\,\,& |r|\in (u_0, 1) .
\end{array}
\right.
\end{equation}
Compact formulas for the radii $u_0$ and $u_1$ are unknown, surely they are $B$ and $M$ dependent. 

(iii) In the $\rho_+$+ $\rho_0$ state one has the following:
\begin{equation}
\frac{\rho_1(r)}{\rho(0)} = \left\{
\begin{array}{ccl}
0, & \,\,\,\,& |r|\in(0, r_2), \\
1 - r^2, & \,\,\,\,& |r|\in (r_2, 1) ,
\end{array}
\right.
\end{equation}
and
\begin{equation}
\frac{\rho_0(r)}{\rho(0)} = \left\{
\begin{array}{ccl}
1-r^2, & \,\,\,\,& |r|\in(0, r_2), \\
0, & \,\,\,\,& |r|\in (r_2, 1) .
\end{array}
\right.
\end{equation}
The radius $r_2$ of the $\rho_0$ domain is a solution of the equation
\begin{equation}
r_2^3-3r_2+2\left(1+M/N \right)=0\, ,
\end{equation}
and is derived from the difference $N=M$. There is only one real solution such that $r_2\in[0, 1]$.

\section{Derivation of critical magnetic fields}
\label{app:B1B2}

In the $\rho_+$+ 2C + $\rho_0$ state there is the phase separation into three stationary domains of the $\rho_+$, 2C and $\rho_0$ phases, see Fig.~\ref{fig:fig1}a; and similarly for the $\rho_+$+ $\rho_0$ state, but this time stationary domains are of the $\rho_+$ and $\rho_0$ phases, see Fig.~\ref{fig:fig1}c. Based on stability conditions for the coexistence of two phases, it is possible to reach $r$ dependence of critical magnetic fields. Here we follow our previous analysis~\cite{ourPRB} keeping the harmonic trap potential in all steps. One starts with stationary Gross-Pitaevskii equations in the TF limit:
\begin{eqnarray} \label{GPE}
0 &=& \left(V(x) +c_0\rho+AB^2+\gamma-\mu\right)\psi_1 + \nonumber\\
      &&+c_2\left[(\rho_1-\rho_{-1})\psi_1+\rho_0\psi_1+\psi_{-1}^*\psi_0^2\right],\\
0 &=& \left(V(x) +c_0\rho-\mu\right)\psi_0 + \nonumber
      c_2\left[(\rho_1+\rho_{-1})\psi_0+2\psi_0^*\psi_1\psi_{-1}\right],\\
0 &=& \left(V(x) +c_0\rho+AB^2-\gamma-\mu\right)\psi_{-1} + \nonumber\\
      &&+c_2\left[(\rho_{-1}-\rho_1)\psi_{-1}+\rho_0\psi_{-1}+\psi_1^*\psi_0^2\right],\nonumber
\end{eqnarray}
where $\mu$ is the chemical potential and $\gamma$ is a Zeeman-like Lagrange multiplier used to enforce the desired magnetization $M$. Sufficiently deeply inside each domain $\psi_j$ still follow~(\ref{GPE}).

\emph{(i) The first critical magnetic field}. In the $\rho_+$+ 2C+$\rho_0$ state occurs: inside the $\rho_+$ phase we have $\psi_0=\psi_-=0$, 
in 2C phase $\psi_0=0$ and in $\rho_0$ phase $\psi_+=\psi_-=0$. 
The chemical equilibrium between coexisting phases requires equalization of chemical potentials of two phases, whereas the lack of pressure between different phases is accomplished by equalization of its energy densities. The only nontrivial conditions are determined by equilibrium requirements for the coexisting 2C and $\rho_0$ phases, and they are $c_2\rho_0^2 = c_2m_{2C} + c_0 \rho_{2C}$ and $c_0^2\rho_0^2 = (c_0\rho_0+AB^2)^2$, where $\rho_{2C}=\rho_{+}+\rho_{-}$ and $m_{2C}=(\rho_{+}-\rho_{-})$. Their solution with respect to densities in the limit $c_0 \to \infty$ gives equal densities in both phases and $\rho_{2C}=c_2m_{2C}^2/(2 AB^2)$ (or $AB^2=c_2m_{2C}^2/(2\rho_{2C})$ equivalently). The last implies that the magnetization $m_{2C}$ in the 2C phase, which coexists with the $\rho_0$ phase, is proportional to the magnetic field. Since the magnetization is zero in the $\rho_0$ phase, $m_{2C}$ must be greater than the initial magnetization~\footnote{Initially the system is in the ground 2C+$\rho_+$ state.} $m_{\rm 2C+\rho_+}$ which is
\begin{equation}
m_{\rm 2C+\rho_+}(r) = \rho(0) \left\{
\begin{array}{ccc}
1-r_1^2 , & \,\,\,\,& |r|\in(0, r_1), \\
1-r^2, & \,\,\,\,& |r|\in (r_1, 1) .
\end{array}
\right.
\end{equation}
So for any $B>B_1$ with
\begin{equation}\label{eq:b1}
\left( \frac{B_1(r)}{B_0} \right)^2 = \frac{m_{\rm 2C+\rho_+}(r)^2}{2 \rho(r)},
\end{equation}
the coexistent phases $\rho_+$, 2C and $\rho_0$ form the ground state of the system. 

\emph{(ii) The second critical magnetic field.} This time we look at stability of the $\rho_+$+$\rho_0$ state. Inside the $\rho_+$ phase we have $\psi_-,\psi_0=0$, 
and in the $\rho_0$ phase $\psi_+,\psi_-=0$. The chemical equilibrium and the lack of pressure between phases imply $c_0\rho_0=c_0\rho_++AB^2$ and $c_0\rho_0^2 = (c_2+c_0)\rho_+^2$ respectively. In the limit $c_0\to \infty$ one gets the following equilibrium condition:
\begin{equation}\label{eq:b2}
\left(\frac{B_2(r)}{B_0} \right)^2 = \frac{\rho_+(r)}{2 \rho(0)} \, .
\end{equation}

Notice, the same results (\ref{eq:b1}) and (\ref{eq:b2}) can be obtained the local density approximation by treating the solutions of the homogeneous system~\cite{ourPRB}.


\end{document}